# *OrigamiPlot*: An R Package and Shiny Web App Enhanced Visualizations for Multivariate Data

Yiwen Lu[1,2,3] (yiwenlu@sas.upenn.edu), Jiayi Tong[1,2,4] (jiayi.tong@pennmedicine.upenn.edu), Yuqing Lei[1,2] (lyqlei@pennmedicine.upenn.edu), Alex J. Sutton[5] (ajs22@leicester.ac.uk), Haitao Chu[6,7] (chux0051@umn.edu), Lisa D. Levine[8,9] (Lisa.Levine@pennmedicine.upenn.edu), Thomas Lumley[10] (t.lumley@auckland.ac.nz), David A. Asch[9,11] (asch@upenn.edu), Rui Duan[12] (rduan@hsph.harvard.edu), Christopher H. Schmid[13] (christopher_schmid@brown.edu), Yong Chen[1-3,9,14,15] (ychen123@pennmedicine.upenn.edu)

[1]Center for Health AI and Synthesis of Evidence (CHASE), Department of Biostatistics, Epidemiology and Informatics, Perelman School of Medicine, University of Pennsylvania, Philadelphia, PA, USA
[2]Department of Biostatistics, Epidemiology, and Informatics, University of Pennsylvania Perelman School of Medicine, Philadelphia, PA, USA
[3]The Graduate Group in Applied Mathematics and Computational Science, School of Arts and Sciences, University of Pennsylvania, Philadelphia, PA, USA
[4]Department of Biostatistics, Johns Hopkins Bloomberg School of Public Health, Baltimore, MD, USA
[5]Department of Population Health Sciences, University of Leicester, Leicester, UK
[6]Division of Biostatistics and Health Data Science, University of Minnesota, Minneapolis, MN, USA
[7]Statistical Research and Innovation, Global Biometrics and Data Management, Pfizer Inc., New York, NY, USA.
[8]Pregnancy and Perinatal Research Center, Department of Obstetrics & Gynecology, University of Pennsylvania Perelman School of Medicine, Philadelphia, PA, USA
[9]Leonard Davis Institute of Health Economics, Philadelphia, PA, USA
[10]Department of Statistics, Faculty of Science, University of Auckland, Auckland, New Zealand
[11]Division of General Internal Medicine, University of Pennsylvania, Philadelphia, PA, USA
[12]Department of Biostatistics, Harvard T.H. Chan School of Public Health, Boston, MA, USA
[13]Department of Biostatistics, Brown University, Providence, RI, USA
[14]Penn Medicine Center for Evidence-based Practice (CEP), Philadelphia, PA, USA
[15]Penn Institute for Biomedical Informatics (IBI), Philadelphia, PA, USA

## ABSTRACT

We introduce *OrigamiPlot*, an open-source R package and Shiny web application designed to enhance the visualization of multivariate data. This package implements the origami plot, a novel visualization technique proposed by Duan et al. in 2023, which improves upon traditional radar charts by ensuring that the area of the connected region is invariant to the ordering of attributes, addressing a key limitation of radar charts. The software facilitates multivariate decision-making by supporting comparisons across multiple objects and attributes, offering customizable features such as auxiliary axes and weighted attributes for enhanced clarity. Through the R package and user-friendly Shiny interface, researchers can efficiently create and customize plots without requiring extensive programming knowledge. Demonstrated using network meta-analysis as a real-world example, *OrigamiPlot* proves to be a versatile tool for visualizing multivariate data across various fields. This package opens new opportunities for simplifying decision-making processes with complex data.

**Key words:** Data visualization; Multivariate data; Origami plot

## HIGHLIGHT

1. Maintains the intuitive visual presentation of the radar chart with a modification that avoids the radar chart's potentially misleading aspects.
2. Provides open-source R package and Shiny web app for creating multivariate data visualizations
3. Supports multi-attribute decision-making

# 1. INTRODUCTION

If you are in the market for a new car, you might care about the car's price, comfort, reliability, efficiency, luxury, thrill, or status. If you are considering alternative medical treatments you might care about the effectiveness, risk, cost, or impact on your life. Most things we care about are judged in multiple dimensions. Various graphical tools, from simple bar and pie charts to advanced plots like scatter plot matrices, heatmaps, and radar charts, represent multi-attribute data.[1–6] Radar charts, also known as web charts or spider charts, are commonly used to assess two or more objects (e.g., cars, treatments, hospitals) across multiple attributes.[7–12] However, radar charts are fundamentally misleading if the filled areas are used to represent the effects of objects. In particular, the filled areas are not proportional to the effects and vary according to the order in which those attributes are presented.

To address these challenges, in 2023, Duan et al. proposed an innovative plot -- Origami Plot[13], which, due to its visual similarity, will also be referred to as the *Snowflake Plot*. In this paper, we introduce the R package, *OrigamiPlot*, and its accompanying online interactive Shiny web application. The *OrigamiPlot* package not only simplifies the creation of this novel visual representation but also empowers researchers to uncover patterns and insights that are difficult to detect in traditional plots for multi-attribute data. The package includes four functions: `origami_plot`, `origami_plot_pairwise`, `origami_plot_weighted`, and `area_calculation`.

This paper introduces the functions of *OrigamiPlot* through an illustrative network meta-analysis (NMA) example comparing eight types of prostaglandin treatments for cervical ripening in pregnant women. In NMA, the origami plot serves as a powerful tool to depict the ranking of the effects of treatments across multiple outcomes using both direct and indirect treatment comparisons. This visualization enables clinicians and researchers to better navigate complex treatment networks, offering clearer insights into the overall impact of interventions.

Beyond NMA, the origami plot extends its utility across multiple analytic domains. For example, in population-adjusted indirect comparisons (PAIC), the plot simplifies the visualization of adjustments made to individual patient data, which is particularly useful for comparing clinical outcomes or quality-of-care metrics across hospitals. At the individual level, it facilitates the comparison of clinical outcomes or health statuses, supporting personalized treatment plans based on patient-specific responses. Additionally, in comparative effectiveness research, the origami plot integrates temporal and treatment-specific insights, allowing for the analysis of treatment regimens over time, including variations in administration routes and dosages. This multi-faceted approach enhances clinical decision-making at both individual and population levels, helping inform strategies that prioritize impactful patient care.

## 2. REVIEW OF ORIGAMI PLOT

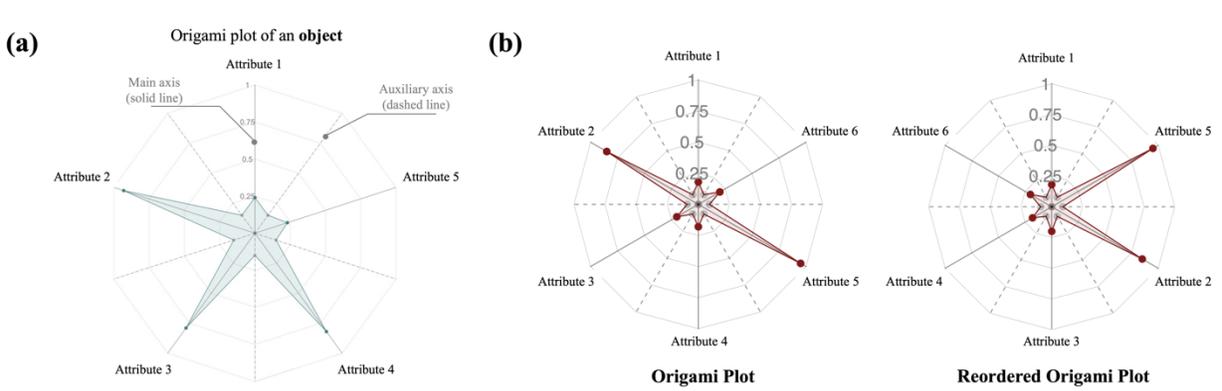

**Figure 1.** (a) An illustration of elements included in an origami plot; (b) A demonstration of the essential invariant area property of origami plot

The origami plot was introduced by Duan et al.[13] as a novel visualization tool. The key elements of an origami plot are shown in **Figure 1(a)**. This plot visualizes multiple attributes of an object simultaneously, with each attribute represented along a different main axis (solid lines). The value of an attribute is represented by a point on each axis, possibly scaled to represent the proportion of the maximum possible value of the attribute. The shortest possible value of the attribute is zero (the center of the plot) and the maximum value is at the outer edge. The plot also features auxiliary axes (dashed lines) with points placed equidistant to the center on each auxiliary axis. The choice of this distance is arbitrary, but it is crucial that the points on the main axes be connected to those on the auxiliary axes. This ensures that the area of the shape formed by connecting the points is invariant to the ordering of the axes. The plot then gives a visual summary of the object's characteristics across multiple attributes.

The design of the origami plot addresses several limitations inherent in radar plots. For example, while maintaining the visual appeal of a radar chart, the origami plot resolves the issue that the area of the connected region changes if the axes are rearranged. In conventional radar charts, varying the order of axes will substantially change the two-dimensional area of the chart. Since those areas are easily misinterpreted as reflecting some overall or combined effect size, radar charts risk creating vastly different interpretations based on an arbitrary choice of axis order. In contrast, the area within Origami plots is invariant to axis order--allowing the area at each point to consistently reflect each individual contribution to overall performance across multiple attributes. In addition to the advantages previously discussed, the origami plot provides benefits like linear scalability of area. In this paper, we will primarily focus on its implementation in the R package. For a detailed exploration of the additional advantages of the origami plot, please refer to Duan et al. (2023)[13].

## 3. EXAMPLE DATASET

To assist users in becoming familiar with the package and its features, *OrigamiPlot* includes a built-in example dataset (*sucra.rda*). This provides data for the network meta-analysis discussed in Duan et al[14]. This example compared eight prostaglandin treatments promoting cervical ripening among women undergoing induction of labor. Five critical outcomes (hereinafter referred to as "five attributes") were examined: need for caesarean section, serious maternal morbidity or death, serious neonatal morbidity or perinatal death, uterine hyperstimulation (when the uterus contracts too frequently or for too long, or does not relax enough between contraction), and vaginal delivery after 24 hours of labor. Duan et al. calculated the Surface Under the Cumulative Ranking (SUCRA) score for the eight treatments[14]. The cumulative ranking curve typically displays sum of the probabilities of each treatment ranking from the best, second, third, and so on, and SUCRA represents the area under this ranking curve. It serves as a metric in network meta-analyses (NMA), ranging from 0 to 1. The closer the value is to 1, the better the treatment ranks overall. SUCRA is also proportional to the mean rank of the treatment. A treatment that is always best (ranked first) has a SUCRA of 1, one that is always worst (ranked last) has a SUCRA of 0, and one that is ranked halfway between first and last would have a SUCRA of 0.5. **Table 1** gives the SUCRA values for the treatments and outcomes from the example dataset *sucra.rda* for the *OrigamiPlot* package. A comprehensive description of the original data used for their NMA analysis and the process to obtain the SUCRA values are provided in the work by Duan et al. (2023)[14].

|  | caesarean | maternal | neonatal | hyperstimulation | vaginal |
|---|---|---|---|---|---|
| Intracervical PGE2 | 0.24 | 0.93 | 0.79 | 0.82 | 0.23 |
| High-dose oral misoprostol | 0.78 | 0.68 | 0.81 | 0.38 | 0.43 |
| Low-dose oral misoprostol | 0.21 | 0.37 | 0.80 | 0.99 | 0.18 |
| Titrated oral misoprostol | 0.93 | 0.58 | 0.44 | 0.54 | 0.82 |
| High-dose vaginal misoprostol | 0.68 | 0.51 | 0.25 | 0.16 | 0.93 |
| Low-dose vaginal misoprostol | 0.69 | 0.58 | 0.23 | 0.33 | 0.79 |
| Vaginal PGE2 | 0.42 | 0.61 | 0.81 | 0.65 | 0.65 |
| Vaginal PGE2 pessary | 0.55 | 0.24 | 0.37 | 0.63 | 0.47 |

**Table 1.** Example dataset *sucra.rda* contained in *OrigamiPlot*. Each row represents one prostaglandin treatment in promoting cervical ripening among women undergoing induction of labor. Each column represents one critical outcome of interest (referred as 'attribute'). Each cell contains the value of SUCRA score for each treatment with respect to each attribute. A SUCRA score closer to 1 is better.

## 4. FUNCTIONS IN *OrigamiPlot* PACKAGE

The *OrigamiPlot* package provides four functions: `origami_plot`, which generates the origami plot for a single object, two variations of the `origami_plot` function: `origami_plot_pairwise`, which creates a pair of origami plots for user-specified objects within a single figure for comparison, and `origami_plot_weighted`, which generates the weighted origami plot, allowing for user-specified adjustments of individual attribute weights to reflect the varying significance of each attribute in a customized manner, and `area_calculation`, which calculates the area of the resulting origami figure. The detailed arguments for each function are provided in **Table 2**.

It is important to note that the *OrigamiPlot* package requires the input data to be arranged as specified (example in **Table 1**): each row must represent one object (e.g., a single treatment, hospital, or individual), and each column should correspond to an attribute. The data values, which are user-defined, can vary widely. For instance, in the example presented in this paper, the SUCRA values range from 0 to 1 for all attributes. The package can create the plots with scale ranging from 0 to 1 by default, but users have the flexibility to customize these values. Note that the *OrigamiPlot* requires a complete input data frame without any missing entries. We recommend using original values of variables when they share the same natural units. In cases where variables are inherently different, converting the original values to comparable units, such as percentiles or rankings, is advised. This ensures that meaningful comparisons can be made and enhances the applicability of the origami plot across various scenarios. **Supplementary Material A** provides details on how the functions process data for use within the package.

## 4.1 `origami_plot` function

`origami_plot` is the main function in the package. It takes a data frame in the required format and constructs an origami plot. The function plots the attributes as solid axes and marks the values corresponding to the attributes on the axes. Additionally, the origami plot also includes the auxiliary axes as dashed lines at equal distances between each neighboring pair of primary axes with auxiliary points pre-specified by the users or automatically generated as half of the minimum value in the input data frame (if the minimum value is 0, the user must manually specify the auxiliary point). Finally, the function connects all the end points on axes in order and obtain a connected region. The area of the connected region generated using `origami_plot` remains consistent regardless of the order of the axes. This is a crucial difference compared to the radar plot for which the area can change when the axes are ordered differently.

**Figure 2** presents eight origami plots corresponding to the eight objects (treatments) in the built-in example cervical ripening dataset (*sucra.rda*). Each plot represents one of the eight objects and visualizes five attributes, corresponding to the five vertices of each polygon. The plot for the Intracervical PGE2 object was generated using the code provided below. The other plots use the same code after changing the object. **Supplementary Material B** describes additional customization options available for plots.

```
origami_plot(sucra, object="Intracervical PGE2")
```

This set of origami plots illustrates the comparative performance of eight prostaglandin treatments across five keys attributes, as introduced in "3. EXAMPLE DATASET". Each axis represents one of these outcomes, with points closer to the edge indicating higher SUCRA scores (approaching 1), which correspond to better rankings for that particular outcome. Treatments with higher scores on multiple axes extend further toward the edge, suggesting a more favorable ranking in those categories. For instance, treatments that minimize risks and promote favorable delivery outcomes are preferred and are visually prominent due to their greater radial reach in the plots. Conversely, axes with shorter extensions indicate areas where a treatment ranks lower, providing a clear visual summary of each treatment's strengths and weaknesses.

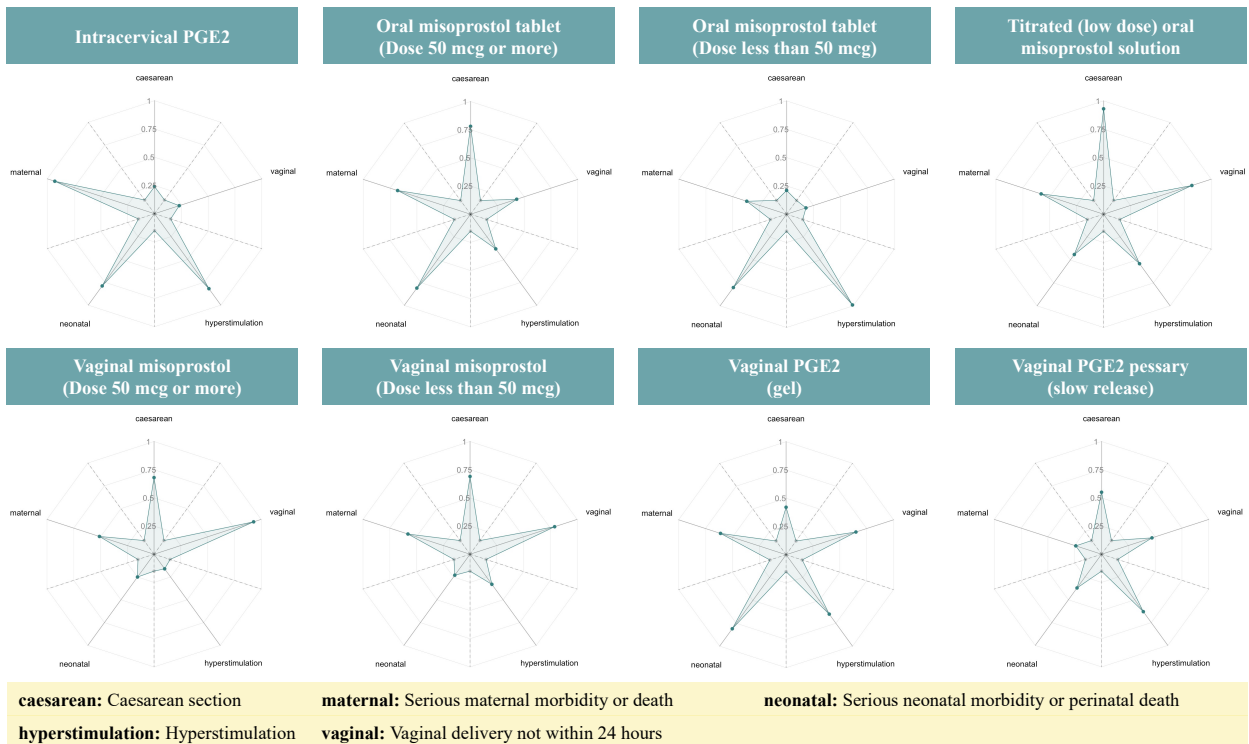

**Figure 2.** Origami plots generated using *origami_plot* with built-in data sets of eight objects and five attributes using sucra.rda.

## 4.2 `origami_plot_pairwise` function

In practice, comparing two objects across multiple attributes is highly valued. To support this, we introduced the `origami_plot_pairwise` function to visualize comparisons between two objects within a single plot. This function supplements the `origami_plot` function which only generates a single plot at a time. The pairwise figure superimposes two plots on top of each other so that they can be compared. Some applications of a pairwise origami plot include comparing two levels of different objects.

- **Hospitals** Two hospitals can be compared on multiple clinical outcomes or quality-of-care metrics in order to illuminate best practices and areas requiring improvement, fostering a competitive spirit aimed at enhancing healthcare quality.
- **Individuals:** Two patients can be compared on clinical outcomes or health status. Such comparisons can be instrumental in personalized medicine, helping to tailor treatment plans based on the differences in patient responses to similar treatments or conditions. This approach can also highlight variance in disease progression and recovery rates, offering deeper insights into patient care.
- **Times**: Outcomes at two different times can be compared within a single patient or hospital in order to track the progress of a patient's condition, the evolution of hospital quality metrics, or the effectiveness of a particular treatment regimen over different time periods. This longitudinal analysis can uncover trends, measure the impact of interventions, and guide future strategies.

- **Treatments**: Two treatments or treatment regimens (route of administration or dose) can be compared at an individual or population level to improve varied clinical outcomes that might be chosen to align with a patient-centered approach exploring which outcomes are the most important for a patient.

As an illustration, we compare outcomes in the example dataset for oral misoprostol and vaginal misoprostol, two treatments that clinicians tend to recommend. Separate comparisons are made for high and low doses. The first pairwise origami plot, comparing high-dose oral misoprostol and high-dose vaginal misoprostol, shown on the left in **Figure 3** is created using the following code.

```
origami_plot_pairwise(sucra,
                     object1=" High-dose oral misoprostol",
                     object2=" High-dose vaginal misoprostol")
```

The pairwise origami plot shown on the right in Figure 3 can be generated in a similar way by adjusting the object argument. The full figure shows that both regimens have a similar risk of cesarean delivery and serious maternal morbidity; however, the vaginal misoprostol has a lower risk of serious neonatal morbidity whereas the oral misoprostol has a lower risk of not achieving a vaginal delivery within 24 hours.

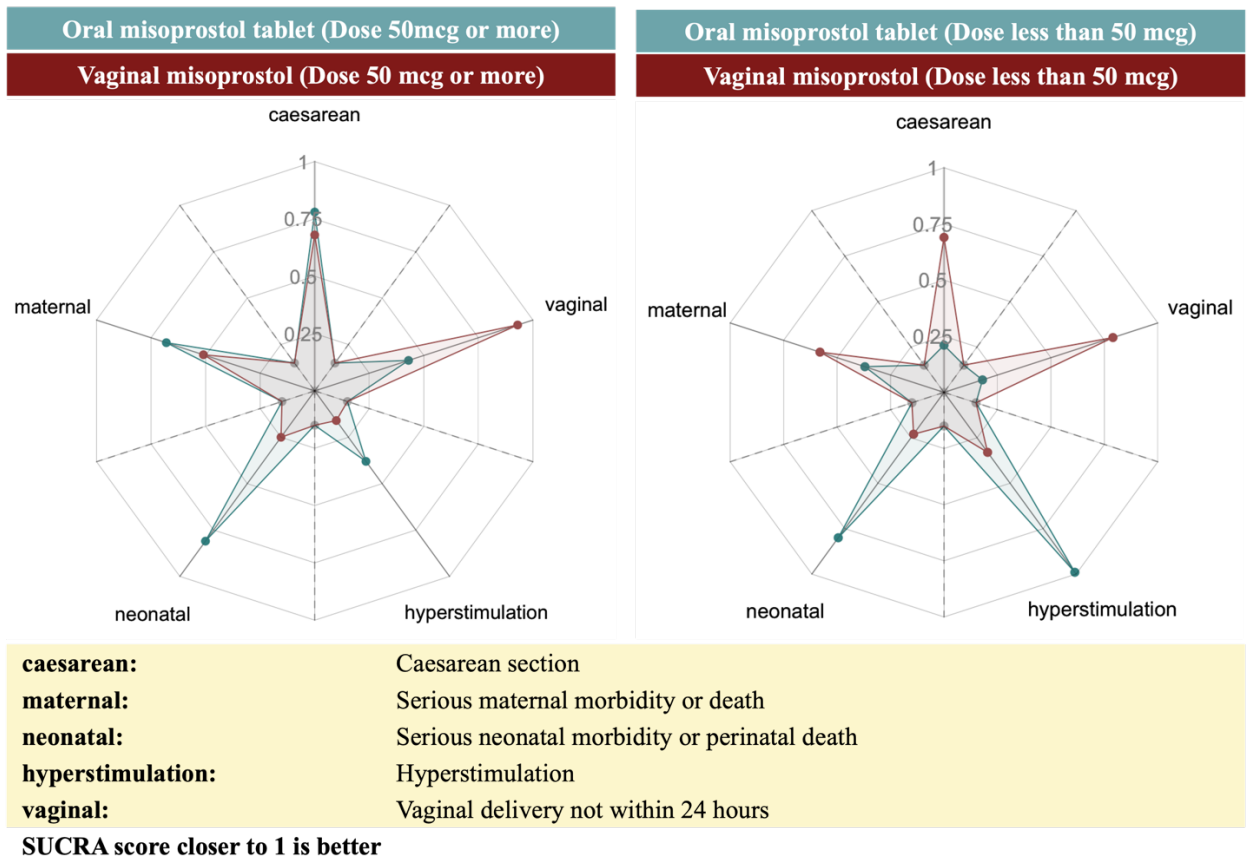

**Figure 3.** Example origami plots generated using `origami_plot_pairwise` with built-in data sets to compare 1. oral misoprostol tablet (dose 50mcg or more) vs. vaginal misoprostol (dose 50 mcg or more); 2. oral misoprostol tablet (dose less than 50 mcg) vs. vaginal misoprostol (dose less than 50 mcg).

## 4.3 `origami_plot_weighted` function

The `origami_plot_weighted` function creates an origami plot with user-specified weights for different attributes. The weighted origami plot is a refined analytical tool that facilitates the adjustment of individual attribute weights to accurately reflect their significance in determining overall performance of the object. For instance, if certain attributes hold greater clinical relevance based on a scientific question, the user can assign higher weights to these attributes relative to others.

This function requires a properly formatted dataset and a positive weight vector that sums to 1, thereby validating the interpretability of the weights. To prevent extreme shrinkage of the origami plot, which could make details difficult to see, we standardize the input weights. We adjust them based on the maximum values to derive a new weight $\frac{w_{input}^k}{\max(\mathbf{W}_{input})}$ for each attribute $k$ where $\mathbf{W}_{input} = \{w_{input}^1, \ldots, w_{input}^K\}$ is the user-specified weight vector. In this way, the attribute with largest input weight has the same value as in the unweighted origami plot.

Here, we demonstrate the weighted origami plot using three objects from the *sucra.rda* dataset, employing synthetic weights. **Figure 4** includes three weighted origami plots. The plot on the left using the oral misoprostol tablet (dose 50mcg or more) was generated using the following code:

```
origami_plot_weighted(sucra, object=" High-dose oral misoprostol",
                      weight = c(0.15,0.25,0.3,0.2,0.1))
```

The other two plots can be generated similarly by changing the object argument.

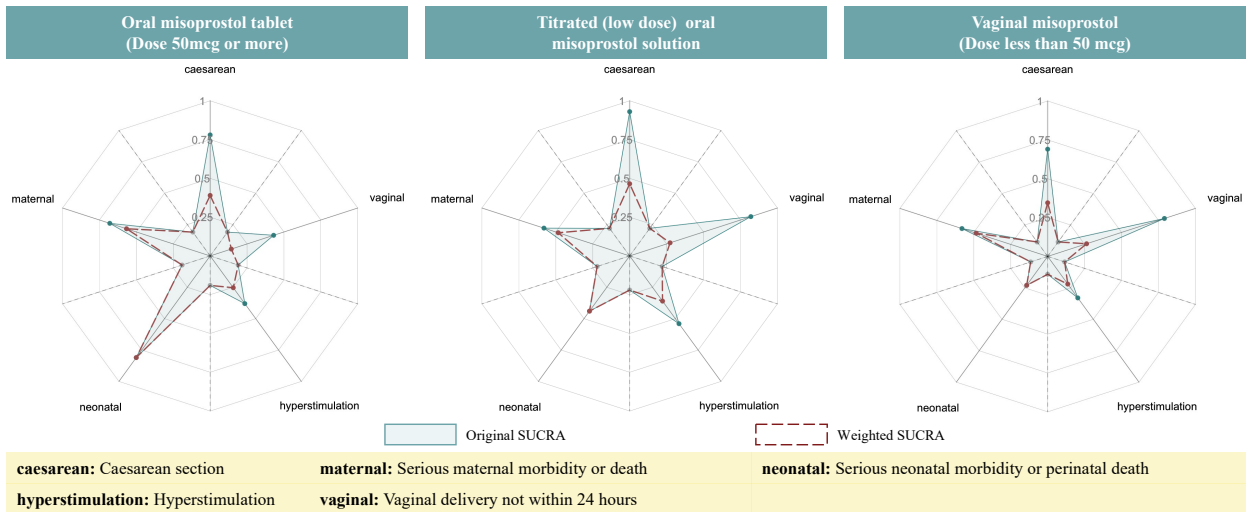

**Figure 4.** Example weighted origami plots generated using *origami_plot_weighted* with synthetic weights and built-in sample data sets oral misoprostol tablet (dose 50mcg or more), titrated (low dose) oral misoprostol solution, and vaginal misoprostol (dose less than 50 mcg).

As shown in **Figure 4**, `origami_plot_weighted` plots two origami plots in a single figure. The first original origami plot, depicted in green, is unweighted and constructed using the SUCRA values on the original scale for each object. In essence, the weights are all equal to 0.20. In contrast, the second origami plot, depicted with a red dashed line, has different weights (0.15, 0.25, 0.3, 0.2, and 0.1) assigned to the attributes. This weighting scheme reflects the differing clinical relevance of each attribute; for instance, serious morbidity or perinatal death is given the highest weight of 0.3 due to its greater clinical significance compared to the others. As a result, this attribute exerts a greater influence on the overall area of the weighted plot. Further details on the area calculation will be discussed in the next section.

## 4.4 `area_calculation` function

The `area_calculation` function is designed to compute the areas of origami plots for objects. This function accepts the same input data format as the visualizing functions above and uses the format presented in **Table 1**. The output generated is the area for each object (i.e., each row of the data input), calculated based on the input values presented on the main axes of the origami plot. By default, the maximum area achievable is 1, assuming the range of input values spans from 0 to 1. A detailed explanation with illustration how to calculate the area can be found in **Supplementary Material C**.

**Function 1:** origami_plot; **Function 2:** origami_plot_pairwise; **Function 3:** origami_plot_weighted; **Function 4:** area_calculation

| Parameter | Function | | | | Description |
|---|---|---|---|---|---|
| | 1 | 2 | 3 | 4 | |
| `df` | X | X | X | X | input dataframe in the required format |
| `object` | X | | X | | the name of the row that user wants to plot |
| `object1` | | X | | | the name of the first row that user wants to plot |
| `object2` | | X | | | the name of the second row that user wants to plot |
| `aux_value` | X | X | X | | Auxiliary point in the graph, default is min(df)/2 |
| `weight` | | | X | | weight of each variable, sum up to 1 |
| `pcol` | X | | X | | color of the line of the (original) polygon, default is rgb(0.2,0.5,0.5,1) |
| `pfcol` | X | | X | | color to fill the area of the (original) polygon, default is rgb(0.2,0.5,0.5,0.1). |
| `pcol 1` | | X | | | color of the line of the first polygon, default is rgb(0.2,0.5,0.5,1) |
| `pfcol 1` | | X | | | color to fill the area of the first polygon, default is rgb(0.2,0.5,0.5,0.1). |
| `pcol 2` | | X | | X | color of the line of the second/weighted polygon, rgb(0.6,0.3,0.3,1) |
| `pfcol 2` | | X | | X | color to fill the area of the second/weighted polygon, default is rgb(0.6,0.3,0.3,0.1). |
| `axistype` | X | X | X | | type of axes. 0:no axis label. 1:center axis label only. 2:around-the-chart label only. 3:both center and around-the-chart labels. Default is 1. |
| `seg` | X | X | X | | number of segments for each axis, default is 4. |
| `pty` | X | X | X | | point symbol, default is 16. 32 means not printing the points. |
| `plty` | X | X | X | | line types for plot data, default is 1 |
| `plwd` | X | X | X | | line widths for plot data, default is 1 |
| `pdensity` | X | X | X | | filling density of polygons, default is NULL |
| `pangle` | X | X | X | | angles of lines used as filling polygons, default is 45 |

| | | | | |
|---|---|---|---|---|
| `cglty` | X | X | X | line type for radar grids, default is 1 |
| `cglwd` | X | X | X | line width for radar grids, default is 0.1 |
| `cglcol` | X | X | X | line color for radar grids, default is #000000 |
| `axislabcol` | X | X | X | color of axis label and numbers, default is #808080 |
| `title` | X | X | X | title of the chart, default is blank |
| `centerzero` | X | X | X | logical. If true, this function draws charts with scaling originated from (0,0). If false, charts originated from (1/segments). Default is TRUE. |
| `vlcex` | X | X | X | font size magnification for vlabels, default is 1 |
| `caxislabels` | X | X | X | center axis labels, default is seq(0,1,by = 0.25) |
| `calcex` | X | X | X | font size magnification for caxislabels, default is NULL |
| `palcex` | X | X | X | font size magnification for paxislabels, default is NULL |

**Table 2.** Description of the arguments available in the functions of the *OrigamiPlot* package.

## 5. R Shiny Web App

Although *OrigamiPlot* was initially designed for use within the command-line R programming environment, we have created a web-based interactive Shiny app to make the tool accessible to users without knowledge in R programming. This app (available at https://origamiplot.shinyapps.io/origami_shinyapp/) is built using Shiny, an R package that enables the creation of interactive web applications. The app offers a graphical user interface (GUI) for the *OrigamiPlot* package, allowing users to engage with its functions without needing to install R or use command-line inputs. Users can upload their data in .csv or .xlsx format to create origami plots with step-by-step guidance and an included example. Single or pairwise origami plots can be generated in the "Origami Plot" tab, while the "Weighted Origami Plot" tab allows users to create weighted plots using specified weights. **Figure 5** displays an example interface, including data uploading section (upper panel), input table (lower left panel), and the origami plot with two bars representing the area of the plot (lower right panel). Detailed instructions and interpretation of the generated figure are provided in both image and video formats within the app.

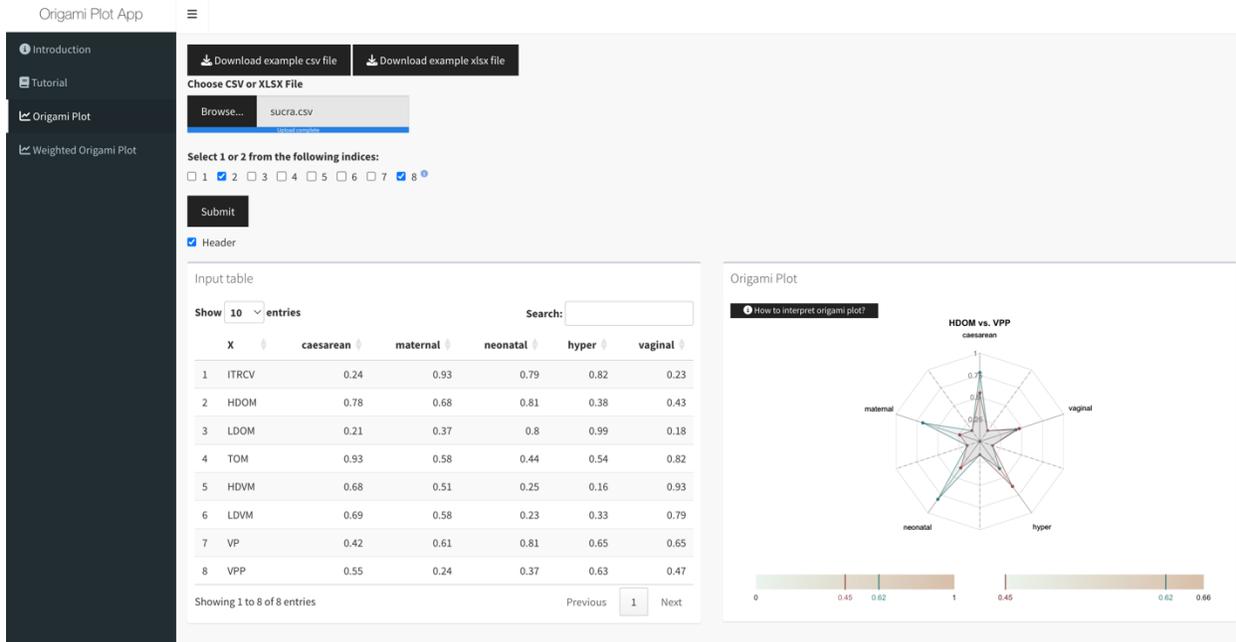

**Figure 5.** Example output of the Shiny web app of *OrigamiPlot*.

## 6. ALIAS OF ORIGAMI PLOT

Given the visual similarity when using the Origami plot to present data with multiple attributes, we have also named our Origami plot the "Snowflake Plot" as an alias. In the R package described above, for user convenience and preference, we have included the following alternative functions based on this alias: `snowflake_plot`, `snowflake_plot_pairwise`, and `snowflake_plot_weighted`. The functionality of these functions remains identical to their origami-based counterparts, ensuring seamless integration and use within analyses. In particular, the input and output parameters are the same as those of the corresponding origami-based functions.

## 7. DISCUSSION

Building on radar charts, we introduce an R package to create origami plots, which retain all the original functionalities of radar charts while mitigating the risk of misinterpreting the connected regions. The origami plot adds auxiliary axes and points to ensure that the area of the connected region remains invariant to the ordering of the axes. This feature allows the origami plot to properly compare multivariate objects across multiple attributes. Additionally, the pairwise origami plot facilitates comparing pairs of objects, thereby augmenting the utility of the proposed origami plot for informed decision-making in contexts that involve multiple attributes. The weighted origami plot enables users to customize attribute weights, allowing for a refined representation of the relative significance of each attribute in a customized and precise manner. In summary, the origami plot is a versatile and informative method for visualizing multivariate data.

The origami plot is particularly useful in comparative effectiveness research for visualizing treatment outcomes, profiling disease risks in epidemiology, comparing treatment effects, and helping identify optimal treatments within broader evidence networks. Looking ahead, the plot's flexibility presents numerous opportunities for expansion. Beyond its current applications, it could be adapted to other fields such as environmental science, where it might visualize complex ecological datasets and capture relationships across multiple environmental variables in an intuitive format. In marketing, the origami plot could profile complex consumer behavior patterns, assisting businesses in understanding multidimensional customer attributes and preferences. In education, it could monitor student performance, compare outcomes across various learning metrics, and pinpoint areas for improvement.

## ACKNOWLEDGEMENT

This work was partially supported in part by National Institutes of Health (1R01LM014344, 1R01AG077820, R01LM012607, R01AI130460, R01AG073435, R01GM148494, R56AG074604, R01LM013519, R56AG069880, U01TR003709, RF1AG077820, R21AI167418, R21EY034179). This work was supported partially through Patient-Centered Outcomes Research Institute (PCORI) Project Program Awards (ME-2019C3-18315 and ME-2018C3-14899). All statements in this report, including its findings and conclusions, are solely those of the authors and do not necessarily represent the views of the Patient-Centered Outcomes Research Institute (PCORI), its Board of Governors or Methodology Committee. This study was partially supported by the National Institute for Health and Care Research (NIHR) Applied Research Collaboration East Midlands (ARC EM) and Leicester NIHR Biomedical Research Centre (BRC). The views expressed are those of the author(s) and not necessarily those of the NIHR or the Department of Health and Social Care.

## CONFLICT OF INTEREST

The author reported no conflict of interest.

## DATA AVAILABILITY STATEMENT

The software and data presented in this paper are freely available on GitHub: https://github.com/Penncil/Origami_Software